\documentstyle[12pt]{article}

\title{Time is life \\{\small A non-specialist's comment on 
extraterrestrial life}}

\author{D.S.L. Soares \\ Departamento de F\'{\i}sica, ICEx, UFMG 
--- C.P. 702 \\ 30161-970,  Belo Horizonte --- Brazil} 

\date{\today}

\begin{document}
\maketitle

\begin{abstract}
The affirmative statement of the existence of 
extraterrestrial life is tentatively raised to the status of a {\it principle}.
Accordingly, Fermi's question is answered and the anthropic principle 
is shown to be falsifiable. The time-scale for the development of 
life on Earth and the age of the universe are the fundamental quantities 
upon which the arguments are framed. 
\end{abstract}

\section{Introduction}
Life is an event which is intrinsically non-deductible from first principles. 
This idea, in a different context, has been claimed and argued by the 
biologist and Nobel laureate Jacques Monod in a 
book published in 1970 ({Monod 1970}).

On the grounds that life can neither be denied nor fully predicted, a further 
step follows, namely, the declaration of the existence of 
{\it extraterrestrial} life as a {\it principle} of Nature. 

The main advantages of a principle for the existence of extraterrestrial
life are: (i) the solution of
paradox-like statements concerning extraterrestrial life (e.g.,
Fermi's question), (ii) the suppression of geo- and anthropocentric ideas,
and (iii) the creation of a logical basis for future theoretical and
experimental investigations. In practice the latter means that one does not
need to justify any scientific project on extraterrestrial life searches 
({for example, Sagan and Drake 1975}) regarding its logical foundations: the 
principle provides (is) the foundation.

The crucial experiment for the origin of life has not to be done; it was
already done on Earth. It seems fair to believe that given a set of 
yet unknown environmental conditions life is bound to flourish. Examples 
of such a conception, i.e., that life is not a privilege of our local 
environment, are multiple in the literature, from the early incursions by 
Giordano Bruno ({e.g., Gatti 1999}) and Christiaan Huygens ({1798}) through 
the modern ages with Robert Goddard ({see Sagan 1979}), Sagan and Salpeter 
({1976}), and others.

It is worthwhile pointing out that the meaning of life used here is 
definitely not restricted to carbon-based 
organisms developed upon watery substrates. A broader concept is envisaged, 
which is not new and may be found, for example, in the investigations by 
Sagan and Salpeter ({1976}) of a possible Jovian ecology, or in the 
literary speculations of a living interstellar cloud, by Hoyle ({1957}), 
and of a structured cometary mind, by MacLeod ({2000}) ---  incidentally, 
both likely being fed by some sort of ubiquitous cosmic plankton. 

Asserting the precise meaning of life is otherwise beyond the scope of the 
present note; the reader is referred to the above-mentioned book by Monod 
for a thorough discussion on the definition of a living organism.

\section{Time is life: a principle}
The principle is set up along the following two lines of arguments.
(1) The existence of life on Earth affirms the crucial experiment for the 
existence of life. (2) The universe is empirically found to be at least three 
times as old as life on Earth. The age of the oldest stars in the Milky Way 
is taken as a {\it lower} limit estimate for the age of the universe, and 
the age of the solar system as an {\it upper} limit estimate for the age of 
life on Earth. Both time-scales are observational facts resting on  
well-established scientific studies. The first one, on the physics of 
energy production in stars, and the second one, on the laws of radioactive 
decay applied to meteorites. 

{\it The universe is old enough such that life and a local ecology 
are expected features of any environment}. 

Time is life, that is to say, give it {\it time} and {\it life} is 
the irremediable end product.

\section{Discussion}
In the light of the principle, a number of other topics deserve renewed 
attention. Below, four of them are briefly touched: Fermi's question, the 
anthropic principle, extraterrestrial intelligence and panspermia.

The famous question posed by Enrico Fermi in an informal conversation 
during a lunch at Los Alamos, in the summer of 1950 [see later account by 
Eric M. Jones ({1985})], became central in the discussion of the existence of 
extraterrestrial civilizations ({e.g., Newman and Sagan 1981}). 
{\it ``--- Where is everybody?''}, asked Fermi, talking about extraterrestrial 
life. The answer to Fermi's question is plain and uninteresting: 
{\it ``--- They are where they belong to''}. Yet they are, states the 
extraterrestrial life principle.

The anthropic principle ({see, for example, Barrow and Tipler 1988}), 
which certainly with justice should be 
dub\-bed {\it the masterpiece of human arrogance}, is thus
irrelevant since the human ecology is but one amongst many.

The search for extraterrestrial intelligence (e.g., the SETI project, 
see {\tt http://seti.planetary.org/}) is strengthened by the 
principle. But an eventual absence of contact with 
extraterrestrial civilizations should not be confused with their 
non-existence. Establishing contact with alien populations is not a 
prerogative of intelligent life but of a given cultural and social 
characteristic of intelligent life (e.g., mercantilism, in the case of 
mankind, as a driving force for contact between distinct societies on 
Earth in the XV and XVI century). 

Finally, it is important to remark that the acceptance, or the eventual 
empirical verification, of the so-called {\it panspermia paradigm} (see 
electronic links to this and related issues in 
{\tt http://www.panspermia.org/}) makes the extraterrestrial life principle 
obvious.

\vskip 1.5cm

{\noindent \it Acknowledgment --- }  I would like to thank Dr. Andr\'e K.T. 
Assis for comments and suggestions on a earlier version of the manuscript.

\vskip 1.5cm

\clearpage

{\noindent \bf References}

\begin{description}
\item Barrow, J.D. and Tipler, F.J. (1988) The Anthropic Cosmological 
Principle, Oxford University Press, Oxford.
\item Gatti, H. (1999) Giordano Bruno and Renaissance Science, Cornell 
University Press, Ithaca.
\item Hoyle, F. (1957) The Black Cloud, Buccaneer Books, Inc., New York.
\item Huygens, C. (1798) The Celestial Worlds Discovered: Conjectures 
Concerning the Inhabitants, Planets and Productions of the Worlds in the 
Planets, Timothy Childs, London.
\item Jones, E.M. (1985) `Where is Everybody?' An Account of Fermi's 
Question, Los Alamos National Laboratory Report LA-10311-MSF, \\ 
{\tt http://lib-www.lanl.gov/la-pubs/00318938.pdf}.
\item MacLeod, K. (2000) The Oort crowd. Nature, 406, 129.
\item Monod, J. (1970) Le Hasard et la Necessit\'e, Editions 
du Seuil, Paris, chap. II.
\item Newman, W.T. and Sagan, C. (1981) Galactic civilizations: Population 
dynamics and interstellar diffusion. Icarus, 46, 293-327.
\item Sagan, C. (1979) Broca's Brain: Reflections on the Romance of 
Science, Random House, New York, chap. 18.
\item Sagan, C. and Drake, F. (1975) The search for extraterrestrial 
intelligence. Scientific American, 232, 80-89.
\item Sagan, C. and Salpeter, E.E. (1976) Particles, environments, and 
possible ecologies in the Jovian atmosphere. Astrophys. J. Supp., 32, 737-755.
\end{description}

\end{document}